# Invariant Discovery of Features Across Multiple Length Scales: Applications in Microscopy and Autonomous Materials Characterization

*Aditya Raghavan,[1] Utkarsh Pratiush,[1] Mani Valleti,[1] Richard Liu,[1] Reece Emery,[1] Hiroshi Funakubo,[2] Yongtao Liu,[3] Philip Rack,[1] Sergei Kalinin[1,4]*

[1] Department of Materials Science and Engineering, University of Tennessee, Knoxville, Tennessee 37909, United States of America

[2] Department of Material Science and Engineering, Tokyo Institute of Technology, Yokohama 226-8502, Japan

[3] Center for Nanophase Materials Sciences, Oak Ridge National Laboratory, Oak Ridge, Tennessee 37830, United States of America

[4] Physical Sciences Division, Pacific Northwest National Laboratory, Richland, Washington 99354, United States of America

_______________________________________

[1] sergei2@utk.edu

[2] upratius@vols.utk.edu

[3] araghav4@vols.utk.edu

## Abstract

Physical imaging is a foundational characterization method in areas from condensed matter physics and chemistry to astronomy and spans length scales from atomic to universe. Images encapsulate crucial data regarding atomic bonding, materials microstructures, and dynamic phenomena such as microstructural evolution and turbulence, among other phenomena. The challenge lies in effectively extracting and interpreting this information. Variational Autoencoders (VAEs) have emerged as powerful tools for identifying underlying factors of variation in image data, providing a systematic approach to distilling meaningful patterns from complex datasets. However, a significant hurdle in their application is the definition and selection of appropriate descriptors reflecting local structure. Here we introduce the scale-invariant VAE approach (SI-VAE) based on the progressive training of the VAE with the descriptors sampled at different length scales. The SI-VAE allows the discovery of the length scale dependent factors of variation in the system. Here, we illustrate this approach using the ferroelectric domain images and generalize it to the movies of the electron-beam induced phenomena in graphene and topography evolution across combinatorial libraries. This approach can further be used to initialize the decision making in automated experiments including structure-property discovery and can be applied across a broad range of imaging methods. This approach is universal and can be applied to any spatially resolved data including both experimental imaging studies and simulations, and can be particularly useful for exploration of phenomena such as turbulence, scale-invariant transformation fronts, etc.

## I. Introduction

Imaging is a fundamental tool for acquiring spatially resolved information across all length scales, from the atomic details captured by high-resolution microscopy to the expansive views of astronomy. Imaging methods are universally important as they allow for the direct visualization of structures and phenomena that are critical to various scientific inquiries and applications. Additionally, simulation serves as another crucial source of imaging data, complementing observational techniques by generating predictable, detailed visualizations of processes that are difficult to observe directly. Together, these imaging sources cover a comprehensive range of scales and conditions, providing essential data for advancing our understanding of the physical world.

Often, the objective is to decipher the complex patterns that images reveal at multiple length scales. Certain phenomena, like fractals, turbulence,[1,2,3,4,5] and those evident in simulations, display characteristics that are invariant across scales. This consistency presents unique opportunities for pattern recognition and analysis. Conversely, in scenarios like atomically resolved images, studies of hierarchical materials, or imaging biological systems, the structural details can vary significantly between different scales. This variability introduces challenges in identifying and understanding the relationships and transitions between scales, necessitating sophisticated analytical approaches to fully interpret the diverse and evolving patterns observed.

Analyzing features across multiple length scales presents a complex challenge, as traditional methods either focus on a single scale or explore only statistical properties spanning scales. Typically, techniques such as correlation functions and scaling analyses are employed,[6,7] but these methods tend to overlook the intricate real-space details of the structures that underpin the data. Such details are crucial for a thorough understanding of the material or phenomenon being studied. For non-hierarchical images, where structure and patterns do not repeat across scales, there is a pressing need for methodologies that can efficiently identify and characterize simple, yet significant, features. Therefore, developing approaches that can adapt to both hierarchical and non-hierarchical systems is essential for advancing our capability to extract meaningful insights from diverse imaging data.

In this work, we introduce the approach of the invariant discovery of features in images using the rotationally, translationally, and affine transform invariant variational autoencoders and illustrate its applications for microscopy on the atomic and mesoscale levels. We further extend this approach to the exploration of time-dependent data such as dynamic movies of chemical transformations, and the exploration of composition-dependent surface topography in combinatorial libraries. This approach is universal, applicable to any spatially resolved data including both experimental imaging studies and simulations, and can be particularly useful for exploration of phenomena such as turbulence, scale-invariant transformation fronts, etc.

## II.    Prior work

One of the key challenges for the discovery of the relevant structural elements in images of microstructures is the variability even in nominally identical elements. For example, in atomic resolution images of the structurally equivalent elements, i.e. having identical bonding patterns, will be observed at different strain conditions. In this case, the analysis is greatly simplified by the fact that individual building blocks can be readily identified. For systems such as grain boundaries or domain walls, the descriptors can be built based on the a priori known microstructural elements, such as domain wall topology and connectivity or image patches centered on the walls. However, for problems such as analysis of the continuous microstructures of complex alloys or ceramics even the nature of the individual building blocks can be uncertain, leading to the joint problem of discovering the relevant microstructural elements accounting for their variability within the class.

Previously, variational autoencoders (VAEs) have been used to explore spatial structures in complex microstructures. The unique property of VAEs is their capability to disentangle the representations of the data, allowing them to discover the parsimonious factors of variation within the data. While for benchmark ML problems such as MNIST handwritten digit dataset, these factors are handwriting styles, and for physics problems[8] these factors of variation are often related to fundamental physical order parameters. For example, in atomically resolved imaging these can be phase composition, polarization, and strain components. In most applications to date, the local descriptors were chosen to be square image patches. The latter can be sampled over the rectangular grid of points, or selected to be centered on atoms, grain or domain boundaries, or other features of interest. Below, we briefly recap the principles of VAE as applied to image analysis problems.

VAEs are a class of generative models that combine the principles of deep learning and probabilistic graphical modeling. They consist of two primary components: an encoder and a decoder[9]. The encoder maps input data into a latent space, representing each input as a distribution over the latent variables. This representation captures the underlying factors of variation in the data. The decoder, on the other hand, reconstructs the input data from the latent space, thereby enabling the generation of new data points that are similar to the original dataset[10]. The objective of a VAE is not just to minimize the reconstruction error but also to ensure that the latent variables are distributed in a manner that is easy to sample from, typically aiming for a Gaussian distribution. This is achieved through a loss function that combines reconstruction loss with a regularization term, which aligns the learned distribution with the prior distribution.[11,12]

Invariant VAEs extend the basic VAE framework by incorporating invariance to certain transformations in the data, such as rotations. A rotation-invariant VAE [13,14] is designed to ensure that its latent representations do not change when the input data is rotated. This is crucial for tasks where the orientation of the input data should not affect the outcome, such as in image recognition or medical imaging. To achieve rotation invariance, VAE has an extra latent dimension that encodes the extra invariance prior, thereby ensuring that the essential features are captured without being affected by their orientation. Similar invariance can be accommodated with respect to translation in the $X$ and $Y$ directions.[14] In case of VAE with rotation and translation invariances, we will have 5 latent variables - angle encoding($\theta$), two translation *(X, Y)* encoding and classical latent variables($Z_1$, $Z_2$, ..$Z_n$) that disentangle other variations in the data. In this paper, we will use 2D latent space ($Z_1$, $Z_2$) unless otherwise specified.

The fundamental limitation of VAEs as applied to image analysis is the inherent challenge of choosing the right descriptor. It is intuitively clear that choosing a small window size will miss the relevant details of microstructural organization, whereas large windows reflect irrelevant complexity and can lead to representation collapse. Hence, developing a principled approach for establishing the right descriptor level is critical for analyzing the microstructures via the VAE approach.

This is particularly important for automated experiments such as structure-property relationship learning in electron and scanning probe microscopy, in which the selection of the proper descriptor size is critical to the convergence of the algorithm. In active learning settings, the trial-and-error approach is severely limited due to the finite instrument budgets.

## III.    Scale-invariant VAE transform

The problem of the discovery of the relevant features in images has been extensively explored over recent decades. In particular, a broad range of methods are based on Scale Invariant Feature Transform (SIFT), a feature detection algorithm in computer vision designed to detect and describe local features in images. The algorithm is particularly robust to changes in scale, noise, illumination, and rotation[15,16,17,18]. SIFT operates by identifying key points in an image and then computing a descriptor for each key point that is invariant to scale and rotation. These descriptors can then be used to match different images with high precision. The process involves several steps: scale-space extrema detection, keypoint localization, orientation assignment, and keypoint descriptor assignment. The robustness of SIFT makes it suitable for object recognition, image stitching, 3D modeling, and gesture recognition, among other applications.

Here, we extend the VAE based image analysis to explore the structural variability across multiple scales in single images and explore this concept in more complex cases of the video data and imaging data with known factors of variation. Unlike SIFT, we use the prior defined set of keypoints to build descriptors but explore the similarity and structural evolution in the descriptors across multiple scales. This approach can be used to explore the microstructure building blocks, select the keypoints based on optimal descriptors, and analyze the system evolution in time and space. Here, we first discuss the basic approach of scale-invariant VAE (SI-VAE), and then extend it to the more complex data.

| | |
|---|---|
| **Algorithm 1 Scale Invariant VAE** | |
| 1: | **Input:** Image |
| 2: | **Initialization:** window_size's |
| 3: | **for** window in window_size's **do** |
| 4: |     get DCNN coordinates (STEM) or uniform grid coordinates (SPM) |
| 5: |     get patches around coordinates |
| 6: |   **for** epoch = 1, 2, 3, … **do** |
| 7: |       scale the patches |
| 8: |       train VAE on patches |
| 9: |       save $z_{mean}$ in list |
| 10: |       save the latent manifold and distribution |
| 11: |   **end for** |
| 12: | **end for** |
| 13: | Further analysis with $z_{mean}$ - PCA, KDE |

The basic SI-VAE algorithm is illustrated in Table I. Here, the input is an image. The set of keypoints is defined in the image space. These can be defined on the rectangular sampling grid, based on physical objects of interest such as domain walls, atoms, etc., or using classical algorithms such as SIFT. With the keypoints defined, we build the descriptors centered on the keypoints, which can typically be square image patches of size w. To enable the scale invariant transform, the range of w is defined as $[\omega_{min}, \omega_{max}]$. The patches are resized to the size wrest that can be chosen to be equal, smaller, or larger than $\omega_{min}$. The set of renormalized patches of original size $\omega_{min}$ is used to initialize the VAE, yielding the corresponding latent distributions, latent representations, and latent images. The latter is defined as the image where each keypoint is visualized using the latent variable as the corresponding color label.

With the VAE initialized, we select the next value of the w to create the descriptors. The latter are fed into the VAE. The process is repeated until the $\omega_{max}$ is reached. The advantage of this dynamic approach is that the substitution of the training data set will continuously be updating VAE weights allowing to maintain the similarity of the latent representations of the similar elements after the resizing, avoiding the initialization uncertainty in VAE. The evolution of the system can be explored using both the instant model, and by encoding the full data set of all window sizes via final VAE. We refer to corresponding latent distributions and latent images as instant and final representations respectively.

We refer to the VAE state while training as the instant state and the VAE state after training as the final state. The output of the SI-VAE process is the evolution of the latent distributions with the window size, the evolution of the latent representations, and the latent images. For example, for the 2D latent space, the corresponding outputs will be latent distribution $R(Z_1, Z_2, \omega)$, latent representation $A(Z_1, Z_2, \omega)$, and latent images $Z_1(x_i, y_i, \omega)$ and $Z_2(x_i, y_i, \omega)$, where $(x_i, y_i)$ are keypoints. For the rectangular sampling grids, these will be images. These analyses can be constructed both for the instant VAE, allowing to monitor the training of the system, and the final VAE that allows consistent comparison of microstructure evolution across scales.

The resulting data sets can be explored using classical dimensionality reduction methods such as Principal Component Analysis (PCA).[19, 20, 21, 22] To visualize the effect of increasing window size for each keypoint, we map the evolution of features corresponding to different window sizes to a single value using PCA. In this manner, PCA components represent the evolution of the latent variables with window size, and PCA loading defines the weights of these components at each keypoint. This approach can be further extended for more complex scenarios such as dynamic data exploring the time evolution of the system, or evolution of the system over e.g. concentration space in combinatorial libraries. In these cases, the PCA components are defined over joint parameter space, e.g. ($\omega$, t) for the dynamic data.

The SI VAE analysis is illustrated in Section IV for the example of ferroelectric domain pattern analysis. The exploration of the dynamics data and combinatorial libraries is discussed in Sections V and VI respectively.

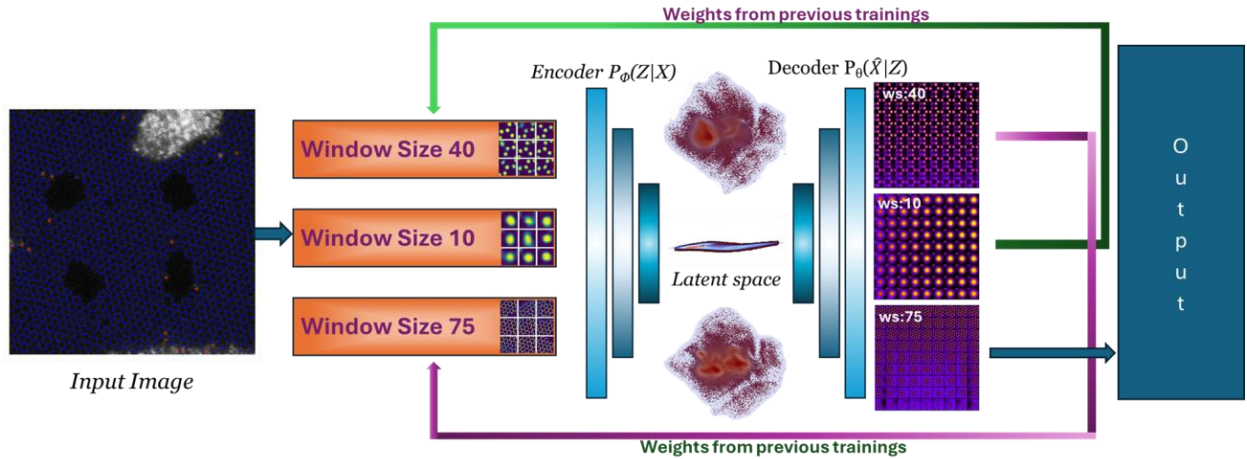

**Figure 1**. Scale Invariant Variational Autoencoder. The input image is sampled over the keypoints (blue carbon atoms in the provided example) to yield image patches of size $\omega$. The image patches at the smallest size and resampled and used to train VAE. Upon convergence, the patches of the next original image size are chosen, and the resultant data is continuously fed to the VAE. The process is repeated until all windows sizes of interest have been analyzed, leading to final VAE model. The evolution of the system can be explored using both the instant model, and by encoding full data set of all window sizes via final VAE.

## IV.    SI-VAE analysis of imaging data

To illustrate the feature discovery using the SI-VAE approach, we choose the example of the multilevel ferroelectric domain images as shown in Figure 2(a). Figure 2(a) shows a vertical PFM amplitude image of a $PbTiO_3$ (PTO) thin film grown on $KTaO_3$ substrates with a $SrRuO_3$ buffer layer[23,24,25], PFM measurement was performed in an Oxford Instruments Asylum Research Cypher AFM using a BudgetSensors ElectriMulti75-G probe. The PFM image reveals both in-plane a domain with polarization vector parallel to the sample surface and out-of-plane c domains with polarization perpendicular to sample surface, as such, the in-plane a domain presents a piezoresponse amplitude close to zero and the out-of-plane domains exhibit a higher piezoresponse. Previously, this material has been explored using active learning-driven autonomous microscopy[26-28] and neural network empowered automated microscopy[29]. Active learning discovered higher polarization dynamics[26] and asymmetric nonlinear behavior across a/c domain walls[27] as well as a broad high nonlinear response near c/c domain walls[27]. Automated piezoresponse spectroscopy measurements empowered by pre-trained neural network revealed alternating high- and low- out-of-plane polarization dynamics near a/c ferroelastic domain walls [29]. These studies indicate that this material represents a well-studied model system.

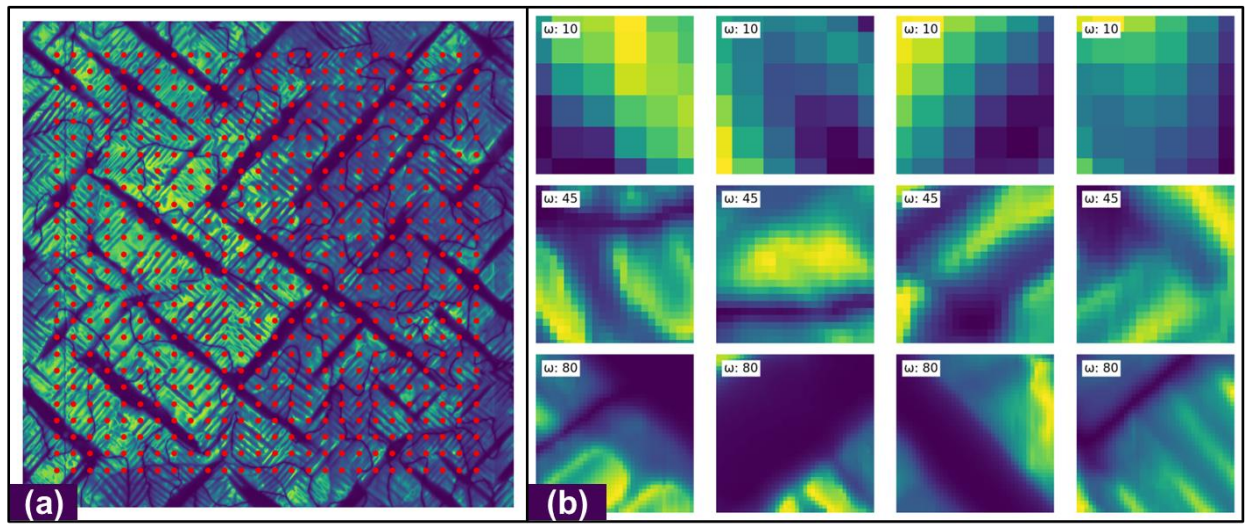

**Figure 2.** **(a)** Image showing the PFM (amplitude) scan with keypoints. **(b)** Patches at different lengthscales, 4 random patches for each window size and we can see that there are more features included as ω increases.

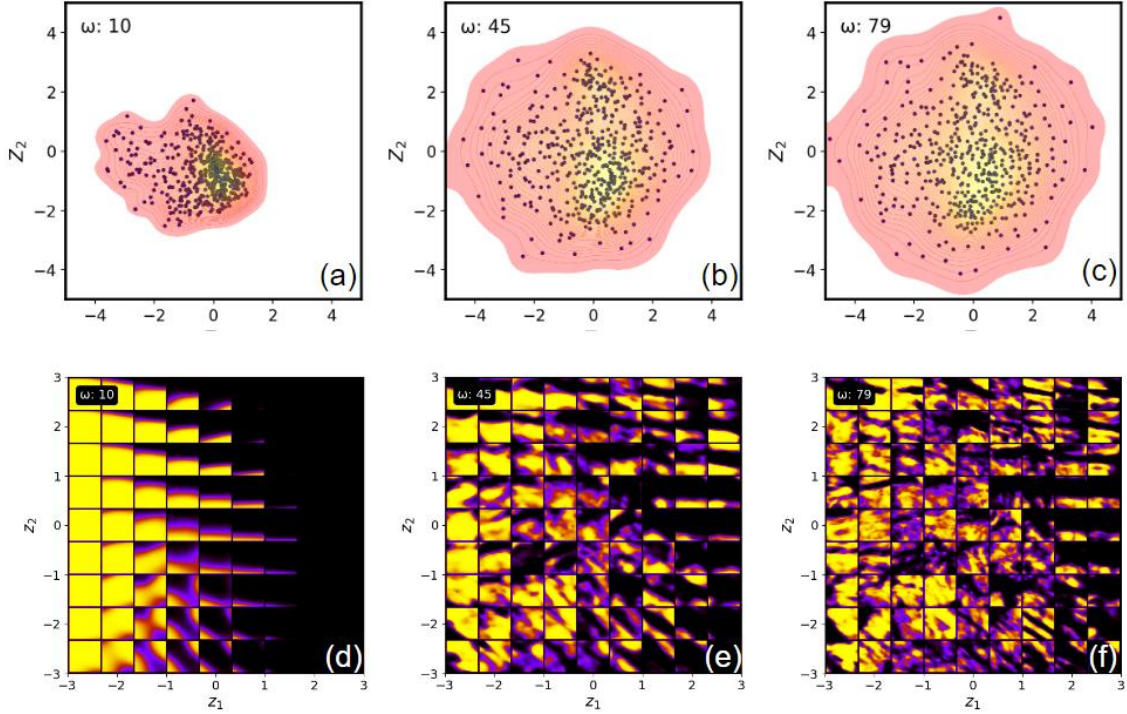

**Figure 3. (a-c)** Latent distributions and **(d-f)** latent representations for the domain images as a function of the window size.

We start with a simple case of a single image to demonstrate how SI-VAE performs and what we can learn from it. The SI-VAE analysis of the ferroelectric domain pattern is illustrated in Figure 3. The evolution of the latent distributions with the window size is shown in Fig. 3(a-c) and illustrates the steady broadening of the latent distribution. This behavior can be expected, since for the small window size (Fig. 2) the patches correspond either to the single domain images or contain a single ferroelectric domain wall. For larger window sizes, the patches start to include more complex microstructural elements such as needle domain termination, periodic patterns, etc. Correspondingly, the complexity of possible objects steadily increases.

This behavior can be further illustrated by latent space representations in Figure 3 (d-f). For small window sizes, the latent space is clearly represented by the unipolar domains, domain walls, and needle domains. Note that while here we have chosen the 9x9 representation of the latent space, the VAE is a generative model and hence the relevant domain structure can be decoded from any latent point. For larger window sizes, the latent representations become more complex, representing the complexity of domain patterns at these length scales.

It is important to note that the VAEs allow for a broad set of invariances including translation and rotation. Hence, the relevant features can be discovered for any rotational and translational states relevant to the sampling grid. Practically, it is important to note that while rotational VAEs generally converge well, translational VAEs work only for relatively small (0 - 0.3) shifts relevant to the patch size. These analysis workflows can be implemented by provided notebooks.

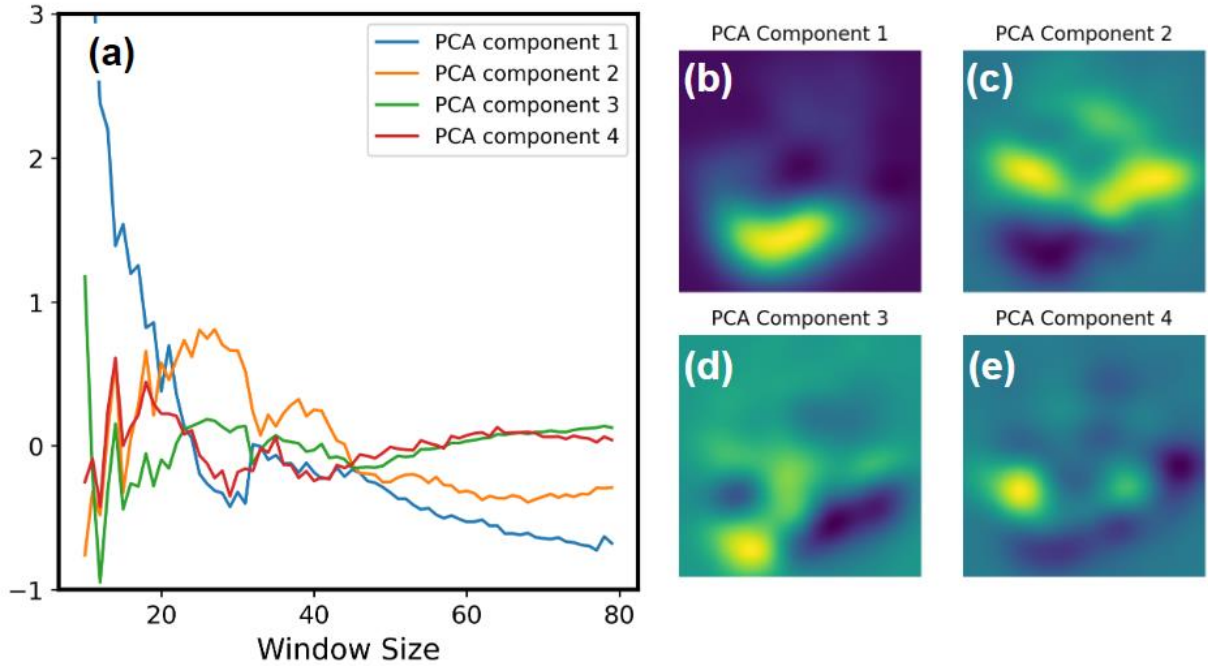

**Figure 4.** PCA analysis of the latent distribution. Shown is **(a)** evolution of the PCA components with the window size and **(b-e)** first 4 PCA components of the kernel density estimate (KDE) of the latent distribution, illustrating the growing complexity in the system with the window size.

To explore the evolution of latent distribution with the window size, we have explored the Principal Component Analysis (PCA). Since the latent embeddings shift with the change in window size, we apply PCA to the Kernel Density Estimate (KDE) of the latent distributions.[30,31,32,33,34] Figure 4(a) illustrates the window size dependence of the PCA components, whereas Figure 4 (b-e) represents the corresponding components. Notably, Component 1 is the most significant at smaller window sizes, exhibiting a rapid decline thereafter. This steep reduction highlights that Component 1 captures a substantial portion of the variance in the initial stages, suggesting that a predominant microstructural feature within the data is most evident when analyzing smaller data subsets. As the window size expands, the prominence of this dominant feature diminishes, likely due to the increasing influence of noise and less correlated variables.

The variance explained by PCA Components 2, 3, and 4 initially rises, then tends to stabilize or decline slightly as the window size continues to expand. This pattern suggests that these components identify behaviors or features that become increasingly evident at larger window sizes. The eventual stabilization indicates that the impact of these features reaches a consistent level, suggesting a plateau in the additional information gained with increased window size. We speculate that this behavior can be further used for the identification of right window sizes for automated experiment, establishing the characteristic lengthscales in the system, and other applications.

We note that this approach can be further extended to find the relevant microstructural elements as the local peaks of the KDE, use the VAE for the outlier detection, and multiple other downstream analysis tasks. However, we defer the investigation of these possibilities for future studies.

## V.    SI-VAE of time-dependent transformations

We further extend the SI-VAE approach for the exploration of time-dependent data, where the features can change jointly across the length scales within the image and with time. Here as a model system, we use the electron-beam induced transformations in graphene. This open data set has been previously used for the time-dependent evolution of the graphene under the action of the electron beam. This data set was acquired as reported elsewhere[14,35,36,37], and illustrates time dependent changes in the graphene monolayer. In the beginning, the material is almost defect free. However, with time, the beam introduces defects associated with the deviation of local structure from the ideal $sp^2$ coordinated carbon atoms. Previously, we have used this data set to illustrate the applications of the VAE to define effective structural order parameters in this system and demonstrate the unsupervised discovery of the chemical building blocks in the materials. Here, we use it to implement the scale-invariant VAE.

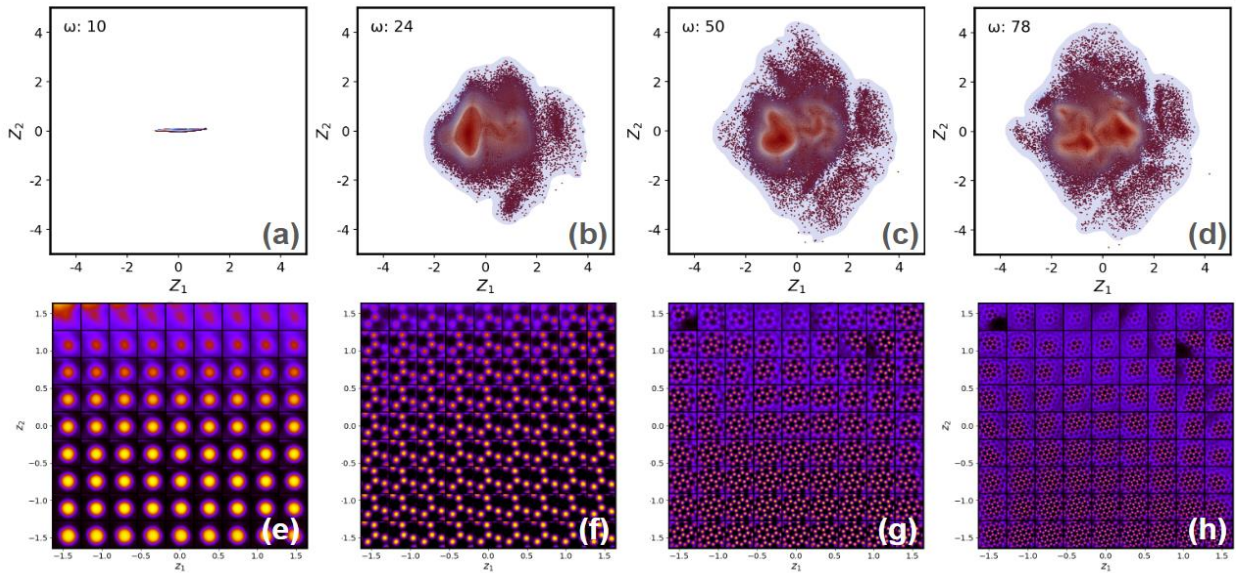

**Figure 5. (a-d)** Window size dependence of the latent distributions and **(e-h)** latent representations for the graphene data set.

Shown in Figure 5 is the evolution of the latent distributions and latent representations as a function of the window size. The windows are taken around atoms detected by pre-trained deep convolution neural network, as described in these references.[38,39] For small window sizes, the latent distribution is effectively 1D, reflecting that at this length scale, the chosen objects are effectively single blobs. This is further visible in the latent representation, for which the latent space is dominated by the blob-like objects. Here, the latent variable is effectively the intensity of the atomic column. Note the emergence of the rod-like objects for high $Z_2$ values; however, these correspond to very low latent densities and represent the (unphysical) extrapolation by the VAE.

On increasing the window size, the structure of the latent distribution becomes progressively more complex, representing the growing structural complexity of the system on larger length scales. For $\omega = 24$, the latent distribution clearly illustrates the discovery of the $sp^2$ coordinated carbon atoms with three nearest neighbors. At this scale, the defects and disorder are presented as atoms with reduced intensity. Note that here we have used the rotationally invariant VAE, for which two types of carbon in ideal graphene are represented as a single pair of latent variables but different latent angles, as further shown in Figure 6 (a).

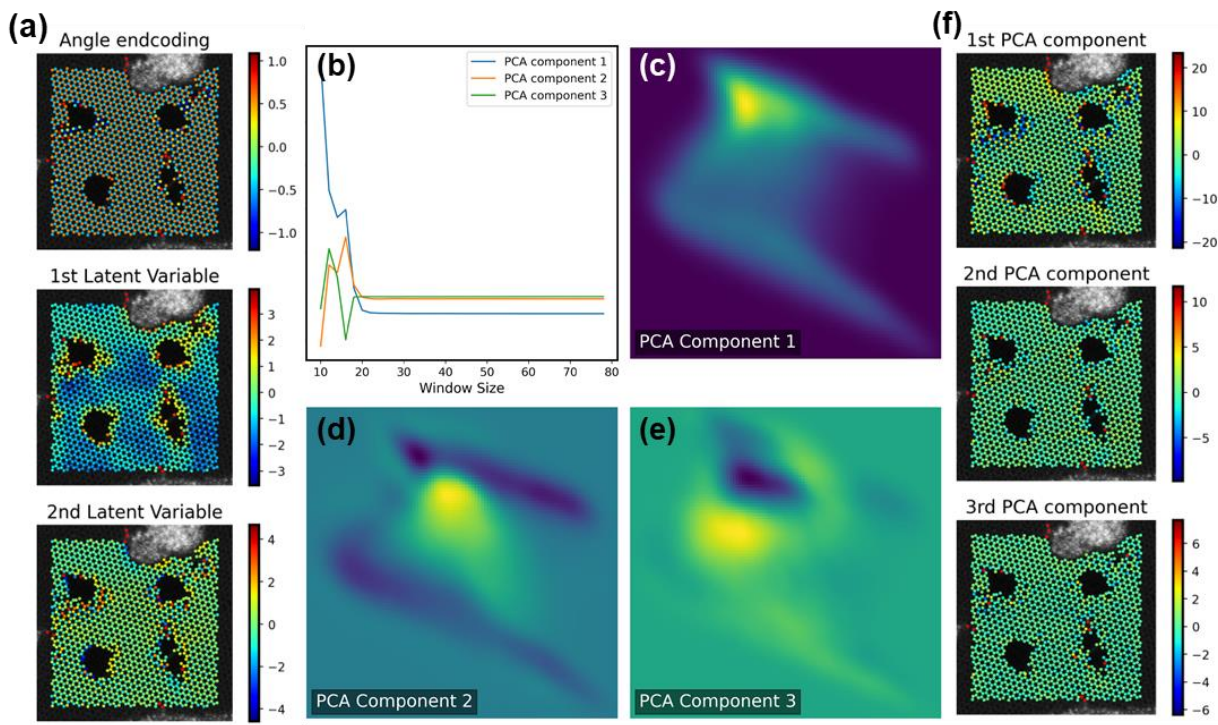

**Figure 6. (a)** Value of latent variables for atoms for window size 70 **(b)** Loading maps of PCA calculated on evolving Latent representation; Visualization of **(c)** PCA component 1 **(d)** PCA component 2, and **(e)** PCA component 3, **(f)** PCA component values for different atoms

Finally, for larger window sizes both the latent distribution and latent representations become more complex, representing the broad structural variability of graphene. $Z_1$ as a function of the window size w can be traced for each atomic unit. Correspondingly, each atom is associated with the latent space trajectory $Z_1(w)$, $Z_2(w)$ that represents the changes in progressively further regions away from the atom. To visualize this behavior, shown in Figure 6 is the PCA of the $Z_1(w)$ trajectories in a single image. Note that for the small window sizes, the distribution is fairly barrow as expected for simple structures at these lengthscales, whereas for large w, the distribution broadens.

The first PCA loading map is reminiscent of the latent image for a sufficiently large window size. Note that the relevant features are generally diffuse, and commensurate with the window size. At the same time, the second and higher loading maps show clear variability of the individual atomic unit behavior, illustrating sharp changes between adjacent atoms. This approach hence allows to separate atoms based on the scale-dependent structure of their neighborhoods.

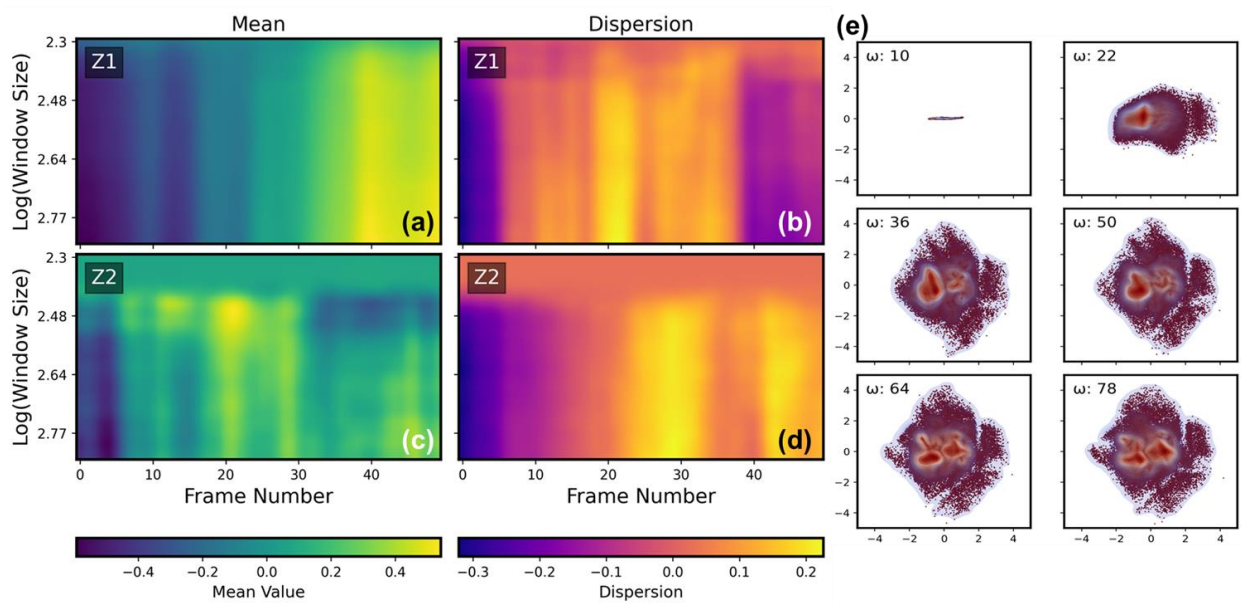

**Figure 7.** Time evolution of Mean and dispersion of (a-b) $Z_1$ and (c-d) $Z_2$. Smoother areas indicate less variability in the latent variables. (e) Latent distributions of different window sizes

Finally, the SI-VAE approach can be further extended to explore the time dynamics of the system. Such analysis can be performed in multiple ways, including the average behavior of the latent variables, the evolution of the latent density peaks corresponding to the most statistically significant microstructural elements, or the width of these peaks corresponding to the structural variability within a single type of object.

Here, we illustrate the latent variables $Z_1$ and $Z_2$, analyzed across multiple frames and varying window sizes to capture their temporal and scaling dynamics. For each variable, we compute both the mean and variance to assess the central tendency and dispersion within specific observational frames. These statistical measures are then normalized by subtracting the mean of each line, effectively highlighting intrinsic fluctuations. To enhance visual clarity and emphasize underlying patterns, the data are smoothed using a Gaussian filter

Shown in Figure 7 is the evolution of the average latent parameters and their distribution as a function of both scale and frame number. Here, for small length scales the changes are minimal, as can be expected given that the primary building block of the system is carbon atoms. However, for larger scales, the time evolution of the latent variable mean and width is clearly visible. Interestingly, the evolution of the $Z_1$ mean and dispersion is almost scale independent, whereas the average of $Z_2$ clearly illustrates three regimes corresponding to the atom-size descriptors, descriptors on the level of the nearest neighbors, and more complex descriptors.

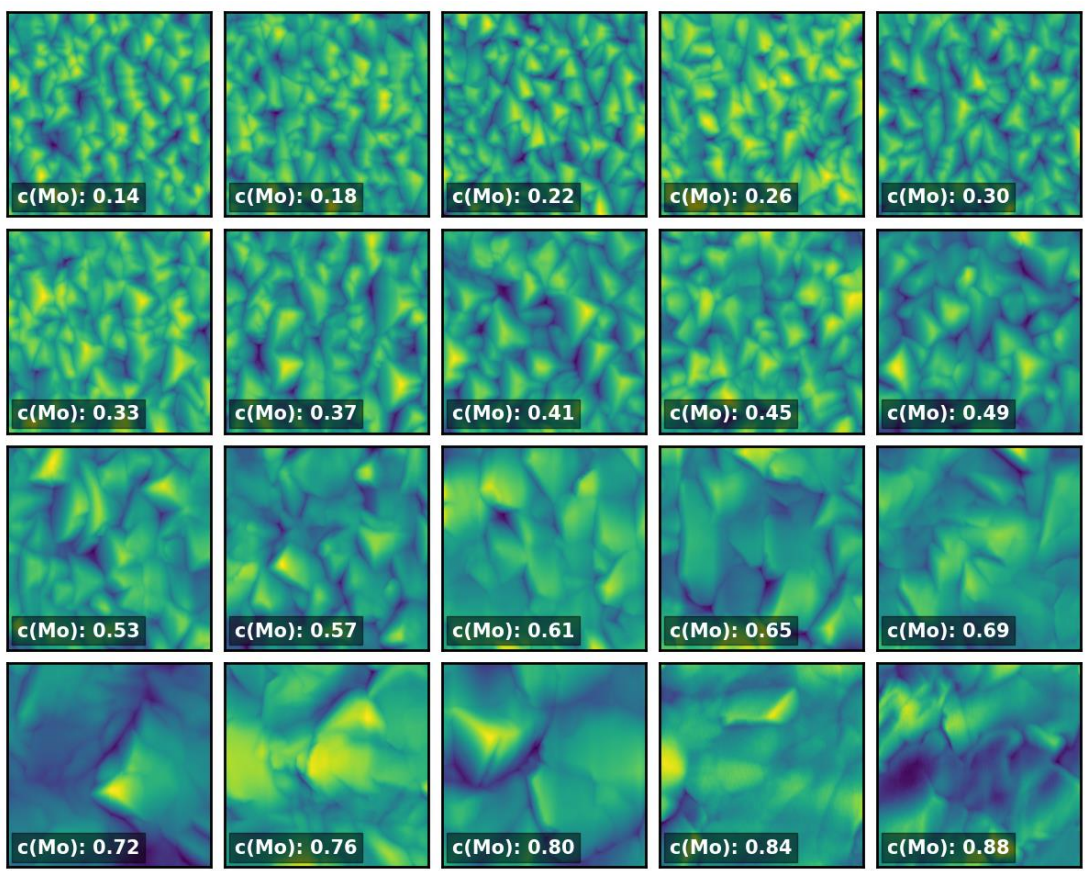

**Figure 8.** The image grid shows the increasing Mo concentration in $Mo_x(CrVTaW)_{1-x}$ system. The growing Mo content in the system influences the grain size which can be seen to vary linearly

## VI.    SI-VAE of combinatorial libraries

We further explore the applicability of the SI-VAE method for the analysis of topography evolution across the combinatorial library. Here we note that surface topography can serve as a proxy signal for the structure zone model[40,41,42] and perhaps the influence that multiple principle elements have on the thin film growth. For example, in the presence of spinodal decomposition or phase separation, the surface will roughen compared to the solid solution regions. Similarly, the concentration-induced differences in the surface mobilities will lead to different surface roughness. However, the length scales at which these behaviors will manifest, and the specific behavior of surface topography are a priori unknown. Here, we explore whether the SI-VAE approach can be used to discover the relevant length scales and traits in concentration-dependent surface behaviors.

As a model system, we use the $Mo_x(CrVTaW)_{1-x}$ system with Mo composition varying across the sample with a range of ($0.14<x<0.88$). This sample was grown via dc magnetron co-sputtering from a Mo and $Cr_{0.25}V_{0.25}Ta_{0.25}Ta_{0.25}$ target at 795 K at 5 *mTorr* Ar sputtering pressure where the Mo and quaternary deposition rates were sputtered at 100 and 200 *W*, respectively, to yield comparable 10 *nm/minute* deposition rates. The goal was to deposit a pseudo binary phase diagram of equiatomic CrVTaW and Mo, where Mo = 0.20 should theoretically be the pentary equiatomic position. Additionally, the Mo melting temperature (2890 *K*) is close to the rule of mixtures estimate of the quaternary CrVTaW melting temperature (2813 *K*)[43]. Based on the sputtering temperature and the homologous substrate temperature (substrate temperature/melting temperature ~ 0.28) a zone *T* microstructure is expected[44].

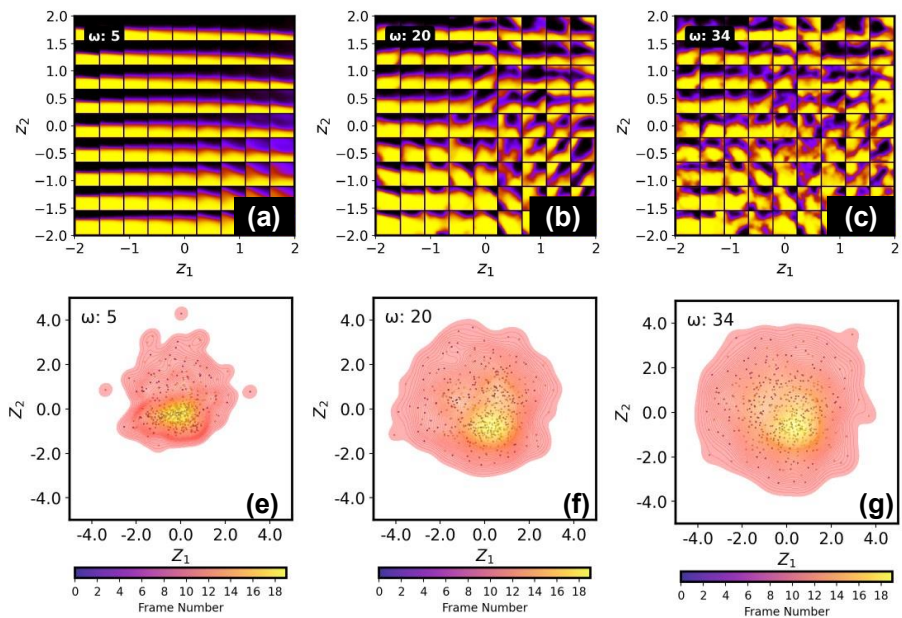

**Figure 9.(a-d)** latent representations for combinatorial library, **(e-h)** Window size dependence of the latent distributions

Subsequent to growth topography images were taken at 20 locations along the direction of varying composition using the automated exploration module in the Oxford instruments Jupiter AFM. This approach has been discussed in ref [45]. The topographic images of the combinatorial library are shown in Figure 8. Here, for the low Mo concentration, the formation of clearly visible triangular features is visible. In the range of $0.14<x<0.57$ range there is a slight monotonic increase in the triangular grains. For higher Mo concentrations the features become larger, and the morphology of the image appreciably changes, as is clearly visible for $c(Mo) > 0.61$. It appears that the zone model shifts from zone 1 to Zone T for Mo > 0.61. While Ludwig et al.[46] observed no significant zone model changes in compositionally complex FCC CoCrFeNi, in this refractory BCC system in the pseudo binary region approaching higher entropy, the compositional complexity does appear to affect the growth. In the future we will correlate diffraction (texture) and physical properties (indentation, resistivity) to the observed topological changes.

The latent representations and latent distributions for the combinatorial library are shown in Figure 9. Here, the evolution of the latent representations and latent distributions is considerably less pronounced than for the ferroelectric domains in Section IV, illustrating the relatively simpler nature of the structural elements visible in Figure 8.

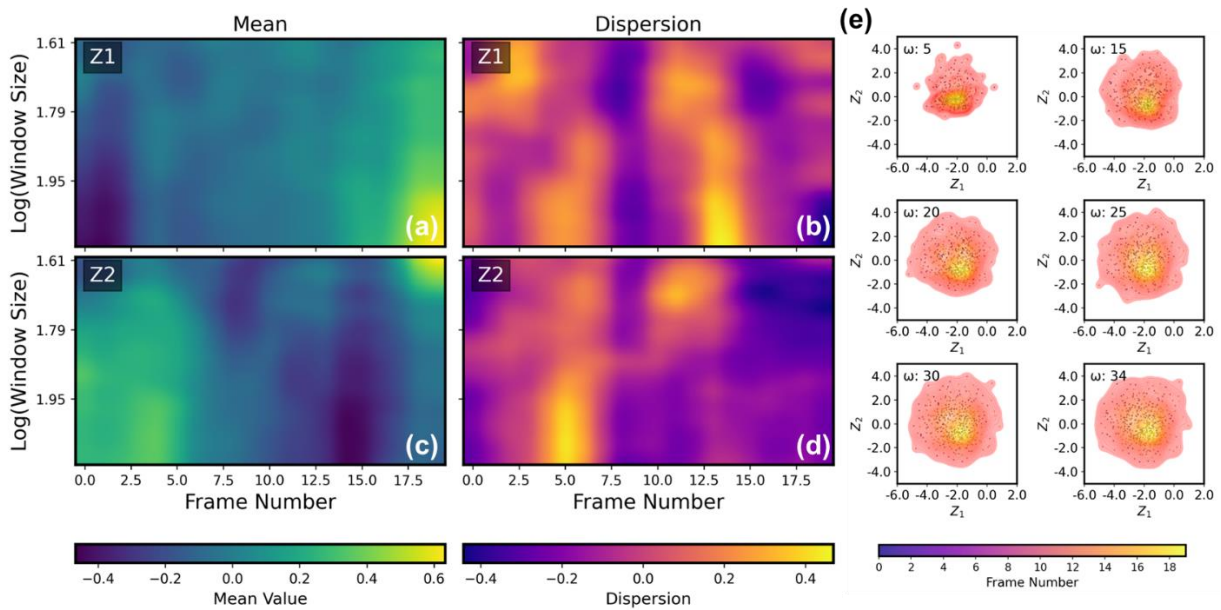

**Figure 10.** Time evolution of Mean and dispersion of **(a-b)** Z1 and **(c-d)** Z2. Smoother areas indicate less variability in the latent variables. **(e)** Latent distributions at different window sizes

The evolution of the SI-VAE along the concentration in the combinatorial library is illustrated in Figure 10(a,c) $Z_1$'s mean values gradually increase, indicating a steady evolution with concentration, while $Z_2$'s mean values decrease, reflecting an inverse relationship. Both variables maintain consistent mean values over different window sizes. The dispersion plots in Figure 10(b, d) reveal regions of heightened variability for both $Z_1$ and $Z_2$, highlighting transient features in the data. Figure 10(e) further illustrates dynamic interactions between $Z_1$ and $Z_2$ across different frames, indicating complex and evolving correlations. This comprehensive analysis underscores the SI-VAE model's capability to capture both steady and transient features, providing valuable insights into the underlying processes of the combinatorial library.

## VII.    Summary

To summarize, here we propose image analysis methods based on the scale-invariant variational autoencoder analysis, SI-VAE. In this, the evolution of system geometry across multiple scales is analyzed by creating and resampling to the same size descriptors with the subsequent VAE analysis. This approach can be applied both for the feature-based descriptors such as individual atomic units, as well as grid-based descriptors. The SI-VAE approach allows to identify the lengthscales at which fundamentally new system geometries emerge. Several visualization approaches of the SI-VAE datasets based on dimensionality reduction are proposed.

This approach is further extended for the analysis of the more complex data sets, including time-dependent movies and images across combinatorial libraries. In these cases, SI-VAE allows to decouple the variability, both across the length scales and control parameter space, and identifies the relevant aspects of system behavior. For example, here we demonstrate the decoding of the chemical evolution in the graphene with time and concentration dependence of the surface morphologies of the combinatorial library of the multicomponent high entropy alloys.

We note that the additional flexibility of the SI-VAE approach can be obtained through the introduction of more complex VAE architectures. For example, the use of the rotationally and translationally invariant VAEs allows us to decouple these factors of variation from the remaining latent variables and explore both their evolution and the simplified latent variables across the parameter spaces. Similarly, we expect that joint and semi-supervised VAEs can be used as a part of the workflow to combine the clustering and representation disentanglement and label propagation tasks respectively.

Finally, we expect that this approach can be beneficial for image analysis in areas such as biology, turbulence, etc., providing information on the scale, translational, and rotational invariance of the structural elements in these systems.

## Acknowledgements

The SI-VAE concept (S.V.K) and implementation (A.R.) supported by the U.S. Department of Energy, Office of Science, Office of Basic Energy Sciences as part of the Energy Frontier Research Centers program: CSSAS-The Center for the Science of Synthesis Across Scales, under award number DE-SC0019288. The AI Tennessee Initiative at UT Knoxville supported the realization of SI-VAE workflows (U.P.) Film growth (H. F.) was supported by by MEXT Program: Data Creation and Utilization Type Material Research and Development Project (JPMXP1122683430). RE and PDR acknowledge their contribution was supported by the National Science Foundation Materials Research Science and Engineering Center program through the UT Knoxville Center for Advanced Materials and Manufacturing (DMR-2309083)

## Data Availablity

The SI-VAE described in this paper are available publicly for use at https://github.com/adityaraghavan98/SI-VAE/tree/main.

# References

[1] R Moss and Roger Moss, University of Oxford, 1992.

[2] L. Danaila, F. Anselmet, and R. A. Antonia, Physics of Fluids **14** (7) (2002/07/01).

[3] Mark Kelly, Wind Energy Science **3** (2) (2018/08/16).

[4] Journal of Wind Engineering and Industrial Aerodynamics **204** (2020/09/01).

[5] Rishita Das and Sharath S. Girimaji, Journal of Fluid Mechanics **941** (2022/06).

[6] Jill Guyonnet, Elisabeth Agoritsas, Sebastian Bustingorry, Thierry Giamarchi, and Patrycja Paruch, Physical review letters **109** (14), 147601 (2012).

[7] Jens Feder, *Fractals*. (Springer Science & Business Media, 2013).

[8] Naoya Takeishi and Alexandros Kalousis, Advances in Neural Information Processing Systems **34**, 14809 (2021).

[9] Diederik P. Kingma and Max Welling, Foundations and Trends® in Machine Learning **12** (4) (2019/06/06).

[10] A Kristiadi, Agustinus Kristiadi's Blog (2020).

[11] Jakub Tomczak and Max Welling, presented at the International conference on artificial intelligence and statistics, 2018 (unpublished).

[12] Mani Valleti, Yongtao Liu, and Sergei Kalinin, arXiv preprint arXiv:2303.18236 (2023).

[13] Tristan Bepler, Ellen Zhong, Kotaro Kelley, Edward Brignole, and Bonnie Berger, Advances in Neural Information Processing Systems **32** (2019).

[14] Sergei V Kalinin, Ondrej Dyck, Stephen Jesse, and Maxim Ziatdinov, Science Advances **7** (17), eabd5084 (2021).

[15] David G. Lowe and David G. Lowe, International Journal of Computer Vision 2004 60:2 **60** (2) (2004/11).

[16] G Lowe, Int. J **2** (91-110), 2 (2004).

[17] Tony Lindeberg, (2012).

[18] E.N. Mortensen, Hongli Deng, and L. Shapiro.

[19] Andrzej Maćkiewicz and Waldemar Ratajczak, Computers & Geosciences **19** (3), 303 (1993).

[20] M. L. Parker, A. C. Fabian, G. Matt, K. I. I. Koljonen, E. Kara, W. Alston, D. J. Walton, A. Marinucci, L. Brenneman, and G. Risaliti, Monthly Notices of the Royal Astronomical Society **447** (1) (2015/02/11).

[21] Jonathon Shlens, arXiv preprint arXiv:1404.1100 (2014).

[22] Ferath Kherif and Adeliya Latypova, in *Machine learning* (Elsevier, 2020), pp. 209.

[23] KATODA Takashi NISHIDE Masamichi, FUNAKUBO Hiroshi, NISHIDA Ken, Journal of the Ceramic Society of Japan **126** (11) (2018-11-01).

[24] Yoshitaka Ehara, Takaaki Nakashima, Daichi Ichinose, Takao Shimizu, Tomoaki Yamada, Ken Nishida, Hiroshi Funakubo, Yoshitaka Ehara, Takaaki Nakashima, Daichi Ichinose, Takao Shimizu, Tomoaki Yamada, Ken Nishida, and Hiroshi Funakubo, Japanese Journal of Applied Physics **59** (SP) (2020-07-23).

[25] Yoshitaka Ehara, Takaaki Nakashima, Daichi Ichinose, Takao Shimizu, Takahisa Shiraishi, Osami Sakata, Tomoaki Yamada, Shintaro Yasui, Ken Nishida, and Hiroshi Funakubo, Applied Physics Letters **121** (26) (2022/12/26).

[26] Yongtao Liu, Kyle P Kelley, Rama K Vasudevan, Hiroshi Funakubo, Maxim A Ziatdinov, and Sergei V Kalinin, Nature Machine Intelligence **4** (4), 341 (2022).

[27] Yongtao Liu, Kyle P. Kelley, Rama K. Vasudevan, Wanlin Zhu, John Hayden, Jon-Paul Maria, Hiroshi Funakubo, Maxim A. Ziatdinov, Susan Trolier-McKinstry, and Sergei V. Kalinin, Small **18** (48) (2022/12/01).

28  Yongtao Liu, Rama K. Vasudevan, Kyle P. Kelley, Hiroshi Funakubo, Maxim Ziatdinov, and Sergei V. Kalinin,  npj Computational Materials 2023 9:1 **9** (1) (2023-03-04).

29  Yongtao Liu, Kyle P. Kelley, Hiroshi Funakubo, Sergei V. Kalinin, and Maxim Ziatdinov, Advanced Science **9** (31) (2022/11/01).

30  Suhas Somnath, Alexei Belianinov, Sergei V. Kalinin, and Stephen Jesse,  Applied Physics Letters **107** (26) (2015/12/28).

31  Shi Na, Liu Xumin, and Guan Yong,  2010 Third International Symposium on Intelligent Information Technology and Security Informatics (2010).

32  Qin Xu, Chris Ding, Jinpei Liu, and Bin Luo,  Pattern Recognition Letters **54**, 50 (2015).

33  Mohiuddin Ahmed, Raihan Seraj, Syed Mohammed Shamsul Islam, Mohiuddin Ahmed, Raihan Seraj, and Syed Mohammed Shamsul Islam,  Electronics 2020, Vol. 9, Page 1295 **9** (8) (2020-08-12).

34  Nancy M. Salem.

35  Maxim Ziatdinov, Ondrej Dyck, Xin Li, Bobby G. Sumpter, Stephen Jesse, Rama K. Vasudevan, and Sergei V. Kalinin,  Science Advances **5** (9) (2019/09).

36  Ondrej Dyck, Songkil Kim, Sergei V. Kalinin, and Stephen Jesse,  Applied Physics Letters **111** (11) (2017/09/11).

37  Ondrej Dyck, Songkil Kim, Sergei V. Kalinin, Stephen Jesse, Ondrej Dyck, Songkil Kim, Sergei V. Kalinin, and Stephen Jesse,  Nano Research 2018 11:12 **11** (12) (2018-07-14).

38  Yongtao Liu, Roger Proksch, Chun Yin Wong, Maxim Ziatdinov, and Sergei V. Kalinin, Advanced Materials **33** (43) (2021/10/01).

39  Maxim Ziatdinov, Ondrej Dyck, Artem Maksov, Xufan Li, Xiahan Sang, Kai Xiao, Raymond R. Unocic, Rama Vasudevan, Stephen Jesse, and Sergei V. Kalinin,  ACS Nano **11** (12) (December 14, 2017).

40  RARR Messier, AP Giri, and RA Roy,  Journal of Vacuum Science & Technology A: Vacuum, Surfaces, and Films **2** (2), 500 (1984).

41  John A Thornton, presented at the Modeling of Optical Thin Films, 1988 (unpublished).

42  Lars Banko, Yury Lysogorskiy, Dario Grochla, Dennis Naujoks, Ralf Drautz, Alfred Ludwig, Lars Banko, Yury Lysogorskiy, Dario Grochla, Dennis Naujoks, Ralf Drautz, and Alfred Ludwig,  Communications Materials 2020 1:1 **1** (1) (2020-03-26).

43  O. El-Atwani, N. Li, M. Li, A. Devaraj, J. K. S. Baldwin, M. M. Schneider, D. Sobieraj, J. S. Wróbel, D. Nguyen-Manh, S. A. Maloy, and E. Martinez,  Science Advances **5** (3) (2019-03).

44  John A. Thornton,  Journal of Vacuum Science & Technology A **4** (6) (1986/11/01).

45  Yu Liu, Utkarsh Pratiush, Jason Bemis, Roger Proksch, Reece Emery, Philip D. Rack, Yu-Chen Liu, Jan-Chi Yang, Stanislav Udovenko, Susan Trolier-McKinstry, and Sergei V. Kalinin, arXiv e-prints (05/2024).

46  Alan Savan, Timo Allermann, Xiao Wang, Dario Grochla, Lars Banko, Yordan Kalchev, Aleksander Kostka, Janine Pfetzing-Micklich, Alfred Ludwig, Alan Savan, Timo Allermann, Xiao Wang, Dario Grochla, Lars Banko, Yordan Kalchev, Aleksander Kostka, Janine Pfetzing-Micklich, and Alfred Ludwig,  Materials 2020, Vol. 13, Page 2113 **13** (9) (2020-05-02).